\documentclass[aps,showpacs,amsmath,amssymb,12pt]{revtex4}
\usepackage{latexsym}
\usepackage{bm}

\def\d{\delta} 
 
\def\m{\mu} 

\def\k{\kappa}
\def\vp{\varphi}
 
\def\t{\tau} 
\def\bx{\bar{\xi}}
\def\bd{\bar{D}}
\def\io{\iota} \def\bio{\bar{\iota}}
 \def\bo{\bar{\omega}}
\def\we{\wedge} 

\def\s{\star} \def\bs{\bar{\star}}
\def\be{\begin{equation}}
\def\ee{\end{equation}}

\begin{document}

\title{Negative Mass Solitons in Gravity} 

\author{Hakan Cebeci}
\email{cebeci@gursey.gov.tr}
\affiliation{Anadolu University, Department of Physics, Yunus Emre
Campus, 26470, Eski{\c s}ehir, Turkey}

\author{{\"O}zg{\"u}r Sar{\i}o\u{g}lu}
\email{sarioglu@metu.edu.tr}
\affiliation{Department of Physics, Faculty of Arts and Sciences,\\
             Middle East Technical University, 06531, Ankara, Turkey}

\author{Bayram Tekin}
\email{btekin@metu.edu.tr}
\affiliation{Department of Physics, Faculty of Arts and Sciences,\\
             Middle East Technical University, 06531, Ankara, Turkey}

\date{\today}

\begin{abstract} 
We first reconstruct the conserved (Abbott-Deser) charges
in the spin connection formalism of gravity for asymptotically (Anti)-de
Sitter spaces, and then compute the masses of the $AdS$ soliton and the 
recently found Eguchi-Hanson solitons in generic odd dimensions, unlike 
the previous result obtained for only five dimensions. These solutions 
have negative masses compared to the global $AdS$ or $AdS/Z_p$ 
spacetimes. As a separate
note, we also compute the masses of the recent even dimensional
Taub-NUT-Reissner-Nordstr{\"o}m metrics. 
\end{abstract}


\maketitle


\section{\label{intro} Introduction}

Energy definition in theories with gravity has been a thorny
issue since the inception of General Relativity. Even though Einstein's 
equation relates local properties of geometry to the local properties of 
matter, when integrated it, nevertheless, 
requires one to express the properties (such as mass) of gravitating 
matter in terms of diffeomorphism invariant geometric quantities. In a 
theory without gravity, such as quantum field theory in flat spacetime,
obtaining the conserved charges, a la Noether,
would be a straightforward task. However, Noether's method, as employed 
by Komar \cite{komar}, leads to certain ambiguities with gravity; such as
assigning different normalization factors - to match the weak field
Newtonian limits - for the mass and angular momenta of asymptotically flat
black hole spacetimes. [Unlike the conserved charges of isolated local
objects in gravity-free theories, one can only talk about the 
\emph{total} energy of a spacetime since gravity cannot be
confined to a region and, as long as diffeomorphism invariance is required, 
one has to talk about the energy of a whole spacetime as opposed to a 
finite domain.]

There are remedies for Komar's method but the problem is that there are
simply `too many' different ones for various spacetimes. There is the
classic work of Arnowitt, Deser and Misner (ADM) \cite{adm} which defines
a Hamiltonian for asymptotically flat spacetimes. The later work of Regge
and Teitelboim \cite{regge} introduces the conserved charges as boundary
terms that are required for a proper variational formulation of the
problem.  For spacetimes which are asymptotically Anti-de-Sitter ($AdS$)
(or asymptotically locally $AdS$), one can find plenty of energy
definitions given \emph{e.g.} by Abbott-Deser \cite{abdes,deser},
Ashtekar-Magnon \cite{ashtekar}, Hawking-Horowitz \cite{hawking}, Aros
\emph{et. al.} \cite{aros}, Cai-Cao \cite{cai}, Henneaux-Teitelboim
\cite{henneaux}, Henningson-Skenderis \cite{henningson} and
Balasubramanian-Kraus \cite{balasubramanian}. [See also Barnich {\it et.
al.} \cite{bar1,bar2} for conserved charges in generic gauge theories,
including gravity.] Unfortunately, each one of these methods have strong
and weak points. A recent detailed comparison of some of these definitions
was nicely carried out by Hollands \emph{et. al.} \cite{hollands}.
Therefore, when an ``interesting" solution for a gravity theory is found,
one wonders about the conserved charges, and more specifically, the mass
of that solution computed by the above methods. Although these methods
frequently agree, one can easily find examples where they don't [See
\emph{e.g.} \cite{chen} in which it was shown that the existence of
long-range scalar fields leads to discrepancies between the Abbott-Deser
and Ashtekar-Magnon definitions for certain spacetimes.]

Recently Clarkson and Mann \cite{clarkson} found new solitons in
cosmological spacetimes that have quite interesting properties: They
resemble the Eguchi-Hanson \cite{eguchi} metrics with $AdS/Z_p$
asymptotics. For the case of a negative cosmological constant, these
solutions have lower energy than the global $AdS/Z_p$ spacetime. 
Clarkson and Mann  computed the energy of the 5-dimensional solution using 
the boundary counterterm 
method of Henningson and Skenderis \cite{henningson}. The authors 
of \cite{clarkson} also claimed that these solutions have the lowest 
energy in their
asymptotic class. In fact, by now, we are used to such novel properties of
$AdS$ spacetimes: Horowitz and Myers \cite{horowitz} provided us with the 
first example of a negative mass soliton, called the ``$AdS$ soliton". These 
negative energy solutions do not cause any instabilities, as in the case 
of a scalar field of negative mass-squared satisfying the 
Breitenlohner-Freedman \cite{breitenlohner} bound. The stability 
of negative mass solitons in the
context of the $AdS$/CFT correspondence is expected since the field theory 
vacuum is stable.

In this paper, we shall compute the masses of both the $AdS$ soliton and
the recently found Eguchi-Hanson (EH) solitons using the Abbott-Deser
\cite{abdes} procedure which can be quite easily generalized to higher
curvature models of gravity \cite{deser}. We would like to stress that,
unlike the boundary counterterm approach which works for a given fixed
dimension, our method applies to generic dimensions and here we will
compute the masses for arbitrary (odd) dimensions. However, before
computing the charges, we will first reconstruct the conserved charges for
cosmological Einstein's theory formulated with the spin connection and the
vielbein instead of the metric. This is a straightforward, yet a tedious
task. Whenever fermionic fields are to be taken into account, such as in
supergravity theories, one has to use the ``first order" spin connection
formulation. We believe that this provides an important motivation as to
why conserved charges in the latter formalism needs to be worked out.

Before we move on to the bulk of the paper, we would like to mention that
observations point out that the Universe might have a small positive
cosmological constant. For this very exciting possibility, in principle,
one would like to study various properties, such as conserved charges,
stability, \emph{etc.} of de Sitter ($dS$) spacetimes as opposed to the
AdS spacetimes. But ironically, most of the recent theoretical progress
(such as the remarkable $AdS$/CFT dictionary) has been on spacetimes with a
negative cosmological constant. Global properties of the latter has
little, if any, resemblance to the former: Therefore, it is not exactly
clear how one would make use of the enormous amount of information gained
in negatively curved spacetimes.  This is a serious challenge but it does
not deter us from studying the $AdS$ spacetimes. In fact, for ``small"
objects (black holes and so on)  that do not change the location of the
cosmological horizon, we want to emphasize that, our formulas will define
mass within the cosmological horizon in de Sitter spacetimes. Moreover
they are also easily modifiable to apply to the higher curvature models,
such as the Gauss-Bonnet theory.

\section{\label{dt} Conserved Charges in Asymptotically $AdS$ Spacetimes}

In this section, we shall first briefly recapitulate the construction
carried out in Abbott-Deser \cite{abdes} and Deser-Tekin \cite{deser}
papers and then redrive the surface integrals for conserved charges in
cosmological Einstein theory using the modern language of differential
forms. As mentioned above, the charge definition that we are about to
present is neither unique nor, in general, in agreement with some other
definitions for all spacetimes. However, we would like to point out that
our definition is quite intuitive and physical: the background spacetime
(the global $AdS$) has zero energy and the asymptotically $AdS$ spacetimes
have energy measured with respect to the background. In some sense, an
observer sitting at the boundary of the spacetime (that is, at the spatial
infinity), sees a black hole as a perturbation to the background
spacetime.  Let us formulate this idea by splitting the metric into a
background plus a perturbation: 
\be 
g_{\mu \nu} \equiv \bar{g}_{\mu \nu} + h_{\mu\nu} \, , 
\ee 
where $g_{\mu \nu}$ is a solution to a certain
gravity theory coupled to matter sources. Note that this theory need not
be Einstein's theory: It could be a complicated higher curvature gravity
model. What we require from this model is that, it either come from a
proper local Lagrangian or it be endowed with the Bianchi identities and
covariant conservation of the matter tensor (or identically, the left hand
side - \emph{i.e.} the geometry part - of the equations of motions). In
what follows ``barred"  quantities refer to the background spacetime that
is a solution to the equations of motion without a source term. We assume
that there are background Killing vectors (to be able to define energy,
one of these vectors has to be a timelike vector everywhere)  
\be
{\bar{\nabla}}_\mu \, \bar{\xi}_\nu^{(a)} + {\bar{\nabla}}_\nu
\bar{\xi}_\mu^{(a)} = 0 \, . \label{killbill} 
\ee 
Having the Killing
equation at our disposal, we can construct partially conserved vector
currents out of the covariantly conserved tensor currents of the
linearized equations. For example, this procedure (worked out in detail in
\cite{deser}) leads to the following conserved charges in cosmological
Einstein theory 
\begin{eqnarray} 
Q^{\mu}(\bar{\xi}) = \frac{1}{4 \,
\Omega_{D-2} \, G_{D}}\int_{\partial M} &dS_i&\Big \{ \bar{\xi}_\nu
\bar{\nabla}^{\mu}h^{i \nu} -\bar{\xi}_\nu \bar{\nabla}^{i}h^{\mu\nu}
+\bar{\xi}^\mu \bar{\nabla}^i h -\bar{\xi}^i \bar{\nabla}^\mu h \nonumber
\\ &&+h^{\mu \nu}\bar{\nabla}^i \bar{\xi}_\nu - h^{i \nu}\bar{\nabla}^\mu
\bar{\xi}_\nu + \bar{\xi}^i \bar{\nabla}_{\nu}h^{\mu \nu} -\bar{\xi}^\mu
\bar{\nabla}_{\nu}h^{i \nu} + h\bar{\nabla}^\mu \bar{\xi}^i \Big \},
\label{ad} 
\end{eqnarray} 
where \( h = h_{\mu \nu} \bar{g}^{\mu \nu} \)
and the gravitational charge has been normalized by the $D$-dimensional
Newton's constant $G_{D}$ and the solid angle of a $(D-2)$-sphere
$S^{D-2}$. Recently, this formula was successfully applied \cite{kanik} to 
the $D$-dimensional Kerr-$AdS$ black holes \cite{gibbons} and a modified
version of it \cite{tekintmg} was used to calculate the charges
of the BTZ black hole \cite{btz} and the charges of the only known
supersymmetric solution to the topologically massive gravity \cite{dereli}
in $D=3$ \cite{serkay}. If there are higher curvature terms present, the 
construction gets modified as worked out in detail in \cite{deser} and 
outlined below.

We now turn on to a detailed computation of the conserved gravitational 
charges formulated with the spin connection and the vielbein. As already 
mentioned in the introduction, such a formulation is forced on us in 
the presence of fermions, for example, in any supergravity theory. [As 
a side note, recall that if the vielbein is assumed to be invertible 
(non-degenerate), then the spin-connection formulation is equivalent 
to the metric formulation.]

Consider a generic gravity theory coupled to a covariantly conserved bounded
matter source which is described by the following `Einstein' equations:
\be
G_{a} + \Lambda \, \s \, e_{a} = \k \, T_{a} \, . \label{eineq}
\ee
Here $G_{a}$ is the `Einstein $(D-1)$-form' of a local {\it 
generic} gravity action, 
$\s$ is the Hodge star operator and $\k$ is
the relevant `coupling constant' of the model under investigation. Suppose now
that the metric tensor $g$
\[ g = \eta_{ab} \, e^{a} \otimes e^{b} , \]
is decomposed such that the `full' orthonormal coframe 1-forms $e^{a}$ can be 
written as the sum of a `background' orthonormal coframe $\bar{e}^{a}$ [which
satisfies (\ref{eineq}) for $T_{a}=0$] plus a `deviation' piece as 
\be
e^{a} \equiv  \bar{e}^{a} + \vp^{a}\,_{b} \, \bar{e}^{b} \, , 
\label{cofram}
\ee
where the 0-forms $\vp^{a}\,_{b}$ are assumed to vanish sufficiently rapidly
at `infinity'. [Note that the decomposition described by (\ref{cofram}) is
always possible given a metric tensor $g$ and a choice for the `background'
coframes $\bar{e}^{a}$, since one can always write 
\( e^{a} = \bar{e}^{a} + \psi^{a}\,_{\m} \, dx^{\m} \) and
\( dx^{\m} = \bar{E}^{\m}\,_{b} \, \bar{e}^{b} , \) for some
0-forms $\psi^{a}\,_{\m}$ and $\bar{E}^{\m}\,_{b}$, which means
\( \vp^{a}\,_{b} = \psi^{a}\,_{\m} \, \bar{E}^{\m}\,_{b} \) in 
(\ref{cofram}).] One can now separate the field equations (\ref{eineq}) 
into a part linear in
$\vp^{a}\,_{b}$ plus all the remaining nonlinear parts so that, one obtains
\[ \bar{G}_{a} (\vp^{b}\,_{c}) = \k \, \t_{a} \, , \]
the `linearized' version of the field equations (\ref{eineq}). Here
$\bar{G}_{a} (\vp^{b}\,_{c})$ is a $(D-1)$-form that involves only terms linear
in the deviation parts $\vp^{b}\,_{c}$ and depends only on the background
coframes $\bar{e}^{a}$ (and, of course, the differential geometric structures 
that they define); the $(D-1)$-form $\t_{a}$ naturally contains all the 
nonlinear terms in $\vp^{b}\,_{c}$ plus the contributions from the original 
matter source $T_{a}$.

It can be shown that due to the background Bianchi identity and the background
gauge invariance, there exists a set, denoted by the index $I$, of 
Killing vectors $\bx_{a}\,^{(I)}$
\be
\bd_{a} \, \bx_{b}\,^{(I)} + \bd_{b} \, \bx_{a}\,^{(I)} = 0 \, , \label{kill}
\ee
for the background geometry described by $\bar{e}^{a}$. Here 
\( \bd_{a} \equiv \bio_{a} \, \bd \); $\bio_{a}$ denotes the interior product
operator with respect to a `background' frame vector that acts on the space of
forms and creates a $(p-1)$-form out of a $p$-form so that, \emph{e.g.} 
\( \bio_{b} \, \bar{e}^{a} = \d_{b}\,^{a} \); $\bd$ denotes the covariant 
derivative operator with respect to the Levi-Civita connection 1-forms
$\bo^{a}\,_{b}$ of the background coframes that satisfy the Cartan structure
equations \( d \bar{e}^{a} + \bo^{a}\,_{b} \we \bar{e}^{b} = 0 \). Since
\( \bd \, \t^{a} = 0 \) by the background Bianchi identity, it readily follows
that one also has
\[ \bd \, (\t^{c} \, \bx_{c}\,^{(I)}) = d (\t^{c} \, \bx_{c}\,^{(I)}) = 0 \, . \]
However, using the fact that the torsion 2-form vanishes, \emph{i.e.} 
\( \bd \, \bar{e}_{a} = 0 \), and defining 
\( \t^{c} = \t^{ca} \, \bs \, \bar{e}_{a} \) for some 0-forms $\t^{ca}$,
one can come up with a conserved density current that leads to the
following conserved Killing charges
\be
Q^{a}(\bx^{(I)}) = \int_{M} \, \tilde{\star} 1 \; \t^{ca} \, \bx_{c}\,^{(I)} =
\int_{\partial M} \, dS_{i} \; q^{ai(I)} \, . \label{genchar}
\ee
Here $M$ is a spatial $(D-1)$-dimensional hypersurface, $\tilde{\star} 1$ is 
the oriented `volume' element of $M$, $\partial M$ denotes its $(D-2)$-dimensional
boundary, we use $dS_{i}$ to denote the corresponding `area' element of $\partial M$,
and $q^{ai(I)}$ is obtained from $\bar{G}_{a} (\vp^{b}\,_{c})$ whose explicit 
form depends on the theory being studied. Here the index $i$ ranges over 
$1, 2, \dots, D-2$ and we have used Stokes' theorem (and the usual 
accompanying assumptions of it) to obtain this final form for the Killing 
charge. [Note that to apply the Stokes' theorem, it is of course necessary 
to write \( \t^{ca} \, \bx_{c}\,^{(I)} = \bd_{c} \, q^{ac(I)} \), which is the 
tricky part but holds for all `physically reasonable' theories that we         
know.]   

Let us be more explicit now and consider the most ``relevant" example 
of the $D$-dimensional cosmological Einstein theory for which the 
vacuum equations read
\be
- \frac{1}{2} \, R^{ab} \, \we \, \s \, e_{abc} + \Lambda \, \s \, e_{c} = 0 
\, .
\label{fullein}
\ee
Here $e_{abc}$ is a shorthand notation for
$e_{a} \we e_{b} \we e_{c}$ and we use analogous expressions
for $e_{ab}$, etc. Vacuum equations 
are solved by a space of constant curvature which satisfies
\begin{eqnarray}
\bar{R}_{abcd} & = & \frac{2 \Lambda}{(D-1)(D-2)} \, ( \eta_{ac} \, \eta_{bd} -
\eta_{ad} \, \eta_{bc} ) \, , \nonumber \\
\bar{R}_{ab} & = & \frac{1}{2} \, \bar{R}_{abcd} \, \bar{e}^{cd} = 
\frac{2 \Lambda}{(D-1)(D-2)} \, \bar{e}_{ab} \, , \label{backgeo} \\
\bar{R} & = & \bio_{b} \, \bio_{a} \, \bar{R}^{ab} = \frac{2 \Lambda D}{D-2} \, . \nonumber
\end{eqnarray}
The `linearization' process of (\ref{fullein}) coupled to a matter 
source in the sense described above involves the use of many nontrivial 
identities and somewhat complicated calculations. We present these 
technical derivations in Appendix A  and proceed with the
explicit form of the first integrand in (\ref{genchar}) which reads
\begin{eqnarray}
\bx_{c} \, \t^{ca} & = & \left( - \bd_{c} \, \bd^{b} \, \vp^{c}\,_{b} +
\bd_{c} \, \bd^{c} \, \vp^{b}\,_{b} + \frac{2 \Lambda}{D-1} \, \vp^{c}\,_{c} 
\right) \bx^{a} - \frac{2 \Lambda}{D-1} \, \bx_{c} \, \vp^{ac} \nonumber \\
& & - \bx_{c} \, \bd^{c} \, \bd^{a} \, \vp^{b}\,_{b}
+ \bx_{c} \, \bd^{c} \, \bd^{b} \, \vp^{a}\,_{b}
- \bx_{c} \, \bd_{b} \, \bd^{b} \, \vp^{ac} + \bx_{c} \, \bd_{b} \, \bd^{a} \, \vp^{bc}
\label{xitau}
\end{eqnarray}
for a given Killing vector $\bx_{c}$ of the `background'. Here, and in what
follows, we suppress the further use of the index $I$ which labels the Killing 
vectors of the background geometry.

The nontrivial task to fulfill now is to put everything on the right hand side
of (\ref{xitau}) in the form $\bd_{c}$(something) and the details of this are given
in appendix \ref{appb}. The outcome of this procedure is 
\begin{eqnarray}
\bx_{c} \, \t^{ca} & = & \bd_{c} \, \left( - \bx^{a} \, \bd^{b} \, \vp^{c}\,_{b} +
\vp^{bc} \, \bd_{b} \, \bx^{a} - \vp^{b}\,_{b} \, \bd^{c} \, \bx^{a} +
\bx^{a} \, \bd^{c} \, \vp^{b}\,_{b} \right. \nonumber \\
& & \left. \qquad - \bx^{c} \, \bd^{a} \, \vp^{b}\,_{b} + \bx^{c} \, \bd^{b} \, \vp^{a}\,_{b} - \bx_{b} \, \bd^{c} \, \vp^{ab} + \vp^{ab} \, \bd^{c} \, \bx_{b} + 
\bx_{b} \, \bd^{a} \, \vp^{cb} \right) \, , \label{charden}
\end{eqnarray}
which explicitly yields the following conserved Killing charge corresponding to
(\ref{genchar})
\begin{eqnarray}
Q^{a}(\bx) & = & \frac{1}{4 \, \Omega_{D-2} \, G_{D}} \, 
\int_{\partial M} \, dS_{i} \,
\left( - \bx^{a} \, \bd^{b} \, \vp^{i}\,_{b} +
\vp^{bi} \, \bd_{b} \, \bx^{a} - \vp^{b}\,_{b} \, \bd^{i} \, \bx^{a} +
\bx^{a} \, \bd^{i} \, \vp^{b}\,_{b} \right. \nonumber \\
& & \left. \qquad \qquad \qquad - \bx^{i} \, \bd^{a} \, \vp^{b}\,_{b} 
+ \bx^{i} \, \bd^{b} \, \vp^{a}\,_{b} - \bx_{b} \, \bd^{i} \, \vp^{ab} + \vp^{ab} \, \bd^{i} \, \bx_{b} +  \bx_{b} \, \bd^{a} \, \vp^{ib} \right) \, 
. \label{char}
\end{eqnarray}
As expected, this is similar in form to the metric formulation 
(\ref{ad}),  but the details were needed to be worked out carefully since 
the spin-connection and the metric formulation are quite distinct in spirit. 
Had we considered a generic higher curvature model in the spin connection 
formulation, the charges would have been modified along the lines of 
\cite{deser}. Here, we will not do that computation, but simply say that, 
for quadratic gravity models, such as the Gauss-Bonnet or any $R^2$ 
theory, a non-trivial factor (depending on the coefficients of the higher 
curvature terms and the cosmological constant) will multiply  
the charge in (\ref{char}). 
 
\section{Computation of the charges for the solitons} 

\subsection{$AdS$ Soliton}

Our first example is the ``$AdS$ Soliton" of Horowitz-Myers 
\cite{horowitz}
\begin{eqnarray}
ds^{2}= \frac{r^2}{\ell^2} \left[ \left( 1 - \frac{r_0^{p+1}}{r^{p+1}}
\right) d\tau^2 + \sum_{i=1}^{p-1} (dx^i)^2 - dt^2 \right] + 
\left( 1 - \frac{r_0^{p+1}}{r^{p+1}} \right)^{-1} \, 
\frac{\ell^2}{r^2} \, dr^2,  
\label{adssoliton}
\end{eqnarray}
which was obtained by the double analytic continuation of a near extremal 
$p$-brane solution. Here $x^i$ ($i = 1, ..., p-1$) and the $t$
variables denote the coordinates on the ``brane"  and $r \ge r_0$. 
To avoid a conical singularity at $r= r_0$, $\tau$ necessarily has a 
period \( \beta = 4 \pi \ell^2 / (r_0 (p+1)) \). 
Its energy was computed in 
\cite{horowitz} using the method of \cite{hawking}. Here we compute the 
energy using the method described so far. The background 
($r_0 = 0$) is the usual globally $AdS$ spacetime in the horospherical 
coordinates, with the timelike Killing vector
\be
\bar{\xi}^\mu = (-1, 0,..., 0). 
\ee
Defining the metric perturbation as outlined above and carrying out 
the integrations, we have 
\be
E = - \frac{V_{D-3} \, \pi}{(D-1) \, \Omega_{D-2} \, G_D} \, 
\frac{r_0^{D-2}}{\ell^{D-2}} \, , 
\ee
where $V_{D-3}$ is the volume of the compact dimensions. Upto trivial 
charge normalizations, our result matches that of \cite{horowitz}, which uses 
the energy definition of Hawking-Horowitz \cite{hawking}.

\subsection{Eguchi-Hanson Solitons}

Recently, Clarkson and Mann \cite{clarkson} found very interesting 
solutions to the \emph{odd} dimensional cosmological (for both signs)  
Einstein equations. These solutions resemble the even dimensional 
Eguchi-Hanson metrics \cite{eguchi} - thus the name Eguchi-Hanson 
solitons - and asymptotically approach $AdS/Z_p$, where $p \ge 3$. As 
shown in 
\cite{clarkson}, these solutions have lower energy compared to the 
global $AdS$ spacetimes (or the global $AdS/Z_p$ spacetimes). 
The energies of 
these solutions (for the case of 5 dimensions) were computed in 
\cite{clarkson} with the help of the boundary counterterm method 
\cite{henningson, balasubramanian}. It is 
important to note that boundary counterterm method needs to be worked 
out for a given fixed dimension. Here, we use the prescription outlined 
in the previous section and find the energy of the EH solitons for 
generic odd dimensions. For a
detailed description of the metrics, we refer the reader to 
\cite{clarkson}. We simply quote their result: the EH soliton reads 
\begin{eqnarray}
ds^{2} &=& -g(r) \, dt^{2} + \left( \frac{2r}{D-1} \right)^{2} \, f(r) \,
\left[ d\psi + \sum_{i=1}^{(D-3)/2} \cos{\theta_{i}} \, d \phi_{i} \right]^{2}  
\notag \\
& & + \frac{dr^{2}}{g(r) \, f(r)} + \frac{r^{2}}{D-1} \, \sum_{i=1}^{(D-3)/2} \, 
\left( d \theta_{i}^{2} + \sin^{2}{\theta_{i}} \, d \phi_{i}^{2} \right) \, ,
\label{EH d-dim}
\end{eqnarray}
and the metric functions are given by 
\begin{equation}
g(r) = 1 \mp \frac{r^{2}}{\ell^{2}} \;\; , \quad 
f(r) = 1 - \left(\frac{a}{r} \right)^{D-1} \, . \label{eh}
\end{equation}
In the $AdS$ case, to remove the string-like singularity at $r=a$, one 
finds that $\psi$ has a period $4 \pi / p$ and there is a constraint on 
the parameter $a$:
\be
a^2 = \ell^2 \, (\frac{p^2}{4} - 1) \, . 
\ee
The background is obtained simply by setting $a=0$ in (\ref{eh}). The 
details of the energy (the charge for $\bar{\xi}^\mu = (-1, 0, 
...,0)$) computation is rather lengthy and not particularly illuminating to 
present here. Instead, we will only write down our result. 
For convenience, we define 
\[ E(\bar{\xi}) \equiv \frac{1}{4 \, \Omega_{D-2} \, G_{D}} 
\, \int_{\partial M} \, dS_{r} \, {\cal{E}}(\bar{\xi}) \, , \]
and only present ${\cal{E}}(\bar{\xi})$, in the $r \to \infty$ limit:
\be
\lim_{r \to \infty} \, 
{\cal{E}}(\bar{\xi}) = - \frac{2 \, a^{D-1}}{\ell^2 \,
(D-1)^{(D-1)/2}} \prod_{i = 1}^{(D-3)/2} \, \sin{\theta_i}.
\ee  
After the angular integrations are carried out, one obtains the energy of 
the EH soliton in generic odd dimensions
\be
E = - \frac{(4 \pi)^{(D-1)/2} \, a^{D-1}}{p \, \ell^2 \, 
(D-1)^{(D-1)/2} \, \Omega_{D-2} \, G_D} \, .
\ee
Specifically, when $D=5$, one finds
\be
E = - \frac{a^4}{4 \, p \, \ell^2 \, G_5} \, .
\ee
We note that this result differs from that of \cite{clarkson} in two 
respects: one of which is a trivial numerical factor that can be attributed 
to normalization of the conserved charges (\ref{char}); the second, and 
the more important one, is the presence of an additive constant which 
is exactly equal to the energy of the $AdS/Z_p$ spacetime. Recall that in the 
formalism we use, the background always has zero energy, unlike the boundary 
counterterm method for which it has a finite energy.   

\section{Conclusions}

In this paper, we have constructed the conserved charges for 
asymptotically $AdS$ spacetimes using the spin connection and the vielbein 
formalism of gravity and then computed the gravitational energies of the
$AdS$ soliton and the recently found EH solitons. For the latter, our method 
provided us with a computation of the masses for generic odd dimensions, 
unlike the boundary counterterm method which was employed only for $D=5$. 
These solitons all have lower energies than the global $AdS$ or the global 
$AdS/Z_p$ spacetimes. 

We would like to stress that the Abbott-Deser \cite{abdes,deser} method, 
which we used here, is a powerful tool that can have a wide range of 
applications.  For example, one can use it to compute the masses of the new 
charged solutions \cite{stelea1,stelea2,stelea3} in $AdS$ spacetimes that 
have non-trivial topology. Here, we only consider two examples that were 
presented in 
\cite{stelea1}. In $D=4$, the ``Taub-NUT-Reissner-Nordstr{\"o}m" solution 
reads \be
ds^2 = -F(r) \, (dt - 2 \, N \, \cos{\theta} \, d\phi)^2 + \frac{dr^2}{F(r)} 
+ (r^2 + N^2) \, (d\theta^2 + \sin^{2}{\theta} \, d \phi^2) \, ,
\ee
where $N$ is the nut charge and 
\be
F(r) = \frac{r^4 + (\ell^2 + 6 N^2) \, r^2 - 2 \, m \, \ell^2 \, r 
- 3 N^4 + \ell^2 \, (q^2 - N^2)}{\ell^2 \, (r^2 + N^2)} \, .
\ee
To find the energy of this solution, the correct background (that has zero 
energy) needs to be carefully chosen. If we naively set $m = q=0$ and the nut 
charge $N = 0$, then the energy of the solution with nonzero $m, q, N$ diverges. 
This is to be expected since $N=0$ solution is \emph{not} in the same 
topological class as that of the $ N\ne 0$ solutions. The background 
has to be chosen as $m=q=0$ but $N \ne 0$ as was shown by Deser-Soldate 
\cite{soldate} in the case of the (asymptotically locally flat) 
Kaluza-Klein monopole. In the light of these arguments, using (\ref{char}) 
one gets 
\be
E = \frac{m}{G_4} \, .
\ee
In $D=6$, the metric, for the details of which we refer to \cite{stelea1}, 
reads
\begin{eqnarray}
ds^2 & = & - F(r) \, (dt - 2 \, N \, \cos{\theta_1} \, d\phi_1 
- 2 \, N \, \cos{\theta_2} \, d\phi_2)^2 + \frac{dr^2}{F(r)} \notag \\
& & + (r^2 + N^2) \, ( d\theta_1^{2} + \sin^{2}{\theta_1} \, d\phi_1^{2} 
+ d\theta_2^{2} + \sin^{2}{\theta_2} \, d\phi_2^{2} ) \, ,
\end{eqnarray}
where now
\begin{eqnarray}
F(r)&=&  \frac{q^2 \, (3r^2 + N^2)}{(r^2 + N^2)^4} \notag \\
& & + \frac{1}{3 \ell^2 (r^2 + N^2)^2} \left[ \ell^2 
(-3 N^4 - 6 m r + 6 N^2 r^2 + r^4) - 15 N^6 + 45 N^4 r^2 
+ 15 N^2 r^4 + 3 r^6 \right] . \nonumber
\end{eqnarray}
Once again the correct background is found by setting $m=q=0$ but $N \ne 0$, 
and the energy is 
\be
E = 12 \, \frac{m}{G_6} \, .
\ee
In both cases, the electric charge $q$ does not appear in the total 
energy just like in the case of ordinary Reissner-Nordstr{\"o}m solution.

\appendix
\section{\label{appa} The derivation of (\ref{xitau})}

In this appendix, we present the technical calculations that lead to 
(\ref{xitau}).

The interior product operator satisfies \( \io_{b} \, e^{a} = \d_{b}\,^{a} \),
which implies that $\bio_{b}$ is related to $\io_{b}$ by
\be
\io_{b} = \bio_{b} - \vp_{b}\,^{c} \, \bio_{c} \, , \label{ibar}
\ee
upto terms first order in $\vp^{a}\,_{b}$. Moreover, substituting 
(\ref{cofram}) in the defining relation of the Hodge operator
\[ \s \, e^{a} \equiv \frac{1}{(D-1)!} \, \epsilon^{a}\,_{bc \dots d} \, e^{b} \, 
\we \, e^{c} \, \we \, \dots \, \we \, e^{d} \equiv \frac{1}{(D-1)!} \, 
\epsilon^{a}\,_{bc \dots d} \, e^{bc \dots d} \, , \]
yields
\[ \s \, e^{a} = \bs \, \bar{e}^{a} + \frac{1}{(D-2)!} \, \epsilon^{a}\,_{bc \dots d} \,
\vp^{b}\,_{p} \, \bar{e}^{pc \dots d} = \bs \, \bar{e}^{a} + \vp^{b}\,_{p} \,
\bar{e}^{p} \, \we \, \bs \, \bar{e}^{a}\,_{b} \]
in terms of the Hodge operator $\bs$ of the `background'. This identity can also
be generalized to
\[ \s \, e^{abc} = \bs \, \bar{e}^{abc} + \vp^{d}\,_{p} \,
\bar{e}^{p} \, \we \, \bs \, \bar{e}^{abc}\,_{d} \]
in a straightforward fashion. When the Cartan structure equations 
\( D e^{a} = d e^{a} + \omega^{a}\,_{b} \we e^{b} = 0 \) are solved for the Levi-Civita
connection 1-forms, one finds
\[ \omega^{a}\,_{b} = \frac{1}{2} \, \left( \io_{b} \, d e^{a} - \io^{a} \, d e_{b}
+ e^{c} ( \io^{a} \, \io_{b} \, d e_{c} ) \right) \, . \]
Since \( d e^{a} = d \bar{e}^{a} + d \vp^{a}\,_{b} \, \we \, \bar{e}^{b}
+ \vp^{a}\,_{b} \, d \bar{e}^{b} \), one obtains (using (\ref{ibar}) and 
(\ref{cofram})) that
\[ \io_{b} \, d e_{c} = \bio_{b} \, d \bar{e}_{c} + (\bio_{b} \, d \vp_{ck}) \,
\bar{e}^{k} - d \vp_{cb} + \vp_{ck} \, (\bio_{b} \, d \bar{e}^{k}) -
\vp_{b}\,^{k} \, (\bio_{k} \, d \bar{e}_{c}) \]
and
\[ \io^{a} \, \io_{b} \, d e_{c} = \bio^{a} \, \bio_{b} \, d \bar{e}_{c} +
\bio_{b} \, d \vp_{c}\,^{a} - \bio^{a} \, d \vp_{cb} + \vp_{ck} \, 
(\bio^{a} \, \bio_{b} \, d \bar{e}^{k}) - \vp_{b}\,^{k} \, (\bio^{a} \, 
\bio_{k} \, d \bar{e}_{c}) - \vp^{ak} \, (\bio_{k} \, \bio_{b} \, d \bar{e}_{c}) \]
up to first order `deviation' terms. Keeping in mind that the `background'
Levi-Civita connection 1-forms satisfy \( \bar{D} \, \bar{e}^{a} = 
d \bar{e}^{a} + \bo^{a}\,_{b} \we \bar{e}^{b} = 0 \), and using the fact that
\( \bd \, \vp^{a}\,_{c} = d \vp^{a}\,_{c} + \bo^{a}\,_{k} \, \vp^{k}\,_{c}
- \bo^{k}\,_{c} \, \vp^{a}\,_{k} \), one thus finds
\[ \omega^{a}\,_{b} = \bo^{a}\,_{b} + \bar{e}^{c} \, [ \bd_{b} \, \vp^{a}\,_{c} -
\bd^{a} \, \vp_{bc} ] \]
after some lengthy but straightforward calculations. Finally the defining expression
\( R_{ab} = d \omega_{ab} + \omega_{ac} \we \omega^{c}\,_{b} \)
yields that the curvature 2-forms of the `background' and the `full' geometry are
related via
\[ R_{ab} = \bar{R}_{ab} - \bar{e}^{c} \, \we \, \bd \, (\bd_{b} \, \vp_{ac} -
\bd_{a} \, \vp_{bc}) \, . \]

When all of these preliminary results are carefully used in (\ref{fullein}),
one obtains the following expression for the `linearized' energy-momentum 
tensor of the
cosmological Einstein theory:
\be
\t_{c} = - \frac{1}{2} \, \bar{R}^{ab} \, \we \, \vp^{d}\,_{p} \, \bar{e}^{p} \,
\we \, \bs \, \bar{e}_{abcd} - \frac{1}{2} \, [\bd \, (\bd^{b} \, \vp^{a}\,_{k} -
\bd^{a} \, \vp^{b}\,_{k})] \, \we \, \bar{e}^{k} \, \we \, \bs \, \bar{e}_{abc}
+ \Lambda \, \vp^{b}\,_{p} \, \bar{e}^{p} \, \we \, \bs \, \bar{e}_{cb} \, . \label{emten}
\ee

Let us now examine the terms in $\t_{c}$ (\ref{emten}) individually. Using
(\ref{backgeo}), the fact that 
\[ \bar{e}^{abp} \, \we \, \bs \, \bar{e}_{abcd} = (D-3) \, \bar{e}^{pa} \, 
\we \, \bs \, \bar{e}_{acd} = (D-3) \, (D-2) \, \bar{e}^{p} \, \we \, \bs \, 
\bar{e}_{cd} \, , \] 
one finds for the first term on the right hand side of (\ref{emten}) that
\[ - \frac{1}{2} \, \bar{R}^{ab} \, \we \, \vp^{d}\,_{p} \, \bar{e}^{p} \,
\we \, \bs \, \bar{e}_{abcd} = - \Lambda \, \left( \frac{D-3}{D-1} \right) 
\, \vp^{b}\,_{p} \, \bar{e}^{p} \, \we \, \bs \, \bar{e}_{cb} \, , \]
which can be added to the last term in (\ref{emten}) to yield
\[ - \frac{1}{2} \, \bar{R}^{ab} \, \we \, \vp^{d}\,_{p} \, \bar{e}^{p} \,
\we \, \bs \, \bar{e}_{abcd} + \Lambda \, \vp^{b}\,_{p} \, \bar{e}^{p} \, \we \, \bs \, \bar{e}_{cb} = \frac{2 \Lambda}{D-1} \, (\vp^{b}\,_{b} \, \bs \, \bar{e}_{c} -
\vp^{b}\,_{c} \, \bs \, \bar{e}_{b}) \, , \]
where we have also used \( \bar{e}^{p} \, \we \, \bs \, \bar{e}_{cb} =
\d^{p}\,_{b} \, \bs \, \bar{e}_{c} - \d^{p}\,_{c} \, \bs \, \bar{e}_{b} \).
The middle term on the right hand side of (\ref{emten}) can be simplified by
first noting that it can be written as 
\begin{eqnarray*} 
- \frac{1}{2} \, [\bd \, (\bd^{b} \, \vp^{a}\,_{k} - \bd^{a} \, 
\vp^{b}\,_{k})] \, \we \, \bar{e}^{k} \, \we \, \bs \, \bar{e}_{abc} & = & 
\frac{1}{4} \, ( \bd_{p} \, \bd^{a} \, \vp^{b}\,_{k} - 
\bd_{p} \, \bd^{b} \, \vp^{a}\,_{k} ) \, 
\bar{e}^{pk} \, \we \, \bs \, \bar{e}_{abc} \\
& & - \frac{1}{4} \, ( \bd_{k} \, \bd^{a} \, \vp^{b}\,_{p} - 
\bd_{k} \, \bd^{b} \, \vp^{a}\,_{p} ) \, 
\bar{e}^{pk} \, \we \, \bs \, \bar{e}_{abc}
\end{eqnarray*}
and this in turn can be further reduced by the fact that
\[ \bar{e}^{pk} \, \we \, \bs \, \bar{e}_{abc} = (\d^{k}\,_{b} \, \d^{p}\,_{a}
- \d^{k}\,_{a} \, \d^{p}\,_{b}) \, \bs \, \bar{e}_{c} + \;\, \mbox{cyclic terms in}
\;\, (a, b, c) \, . \]
Using this, one finally obtains for the middle term on the right hand side 
of (\ref{emten}) that
\begin{eqnarray*}
- \frac{1}{2} \, [\bd \, (\bd^{b} \, \vp^{a}\,_{k} - \bd^{a} \, 
\vp^{b}\,_{k})] \, \we \, \bar{e}^{k} \, \we \, \bs \, \bar{e}_{abc} & = & 
(\bd_{a} \, \bd^{a} \, \vp^{b}\,_{b} - \bd_{a} \, \bd^{b} \, \vp^{a}\,_{b}) \, 
\bs \, \bar{e}_{c} \\ 
& & \hspace{-2cm}
+ (\bd_{c} \, \bd^{b} \, \vp^{a}\,_{b} - \bd_{c} \, \bd^{a} \, \vp^{b}\,_{b} 
+ \bd_{b} \, \bd^{a} \, \vp^{b}\,_{c} - \bd_{b} \, \bd^{b} \, \vp^{a}\,_{c}) 
\, \bs \, \bar{e}_{a} \, . 
\end{eqnarray*}
Combining all of these results finally gives
\begin{eqnarray*}
\t_{c} & = & \eta_{ca} \, \left( - \bd_{p} \, \bd^{b} \, \vp^{p}\,_{b} +
\bd_{p} \, \bd^{p} \, \vp^{b}\,_{b} + \frac{2 \Lambda}{D-1} \, \vp^{p}\,_{p} 
\right) \, \bs \, \bar{e}^{a}  \\ 
& & + \left( - \bd_{c} \, \bd_{a} \, \vp^{b}\,_{b}
+ \bd_{c} \, \bd^{b} \, \vp_{ab} - \bd_{b} \, \bd^{b} \, \vp_{ac} + 
\bd_{b} \, \bd_{a} \, \vp^{b}\,_{c} - \frac{2 \Lambda}{D-1} \, \vp_{ac} \right)
\, \bs \, \bar{e}^{a} 
\end{eqnarray*}
for the `linearized' energy-momentum tensor of the $D$-dimensional
cosmological Einstein theory. Hence (\ref{xitau}) readily follows from this
expression for $\t_{c}$.

\section{\label{appb} The derivation of (\ref{charden})}

In this appendix, we show the details of how (\ref{charden}) is obtained
from (\ref{xitau}). For this purpose, first note the following identities:

Since $\bx_{a}$ is a Killing vector, it immediately follows from the Killing
equation (\ref{kill}) that \( \bd_{a} \, \bx^{a} = 0 \). Moreover, the very
definition of the Riemann tensor implies that 
\[ (\bd_{a} \, \bd_{b} - \bd_{b} \, \bd_{a}) \, \bx_{c} = 
\bar{R}_{abcd} \, \bx^{d} \, . \]
This can be used together with the key property of the Riemann tensor 
[This identity can easily be derived from 
\( D \theta^{a} = R^{a}\,_{b} \, \we \, e^{b} = 0 \) where 
\( \theta^{a} \equiv \bd \, \bar{e}^{a} \) denotes the torsion 2-form.].

\[ \bar{R}_{[abc]d} = 0 \qquad \mbox{and hence} \qquad 
\bar{R}_{[abc]d} \, \bx^{d} = 0, \]
to obtain
\( \bd_{b} \, \bd_{c} \, \bx^{a} = \bar{R}^{a}\,_{cbd} \, \bx^{d} \).
This further simplifies by making use of (\ref{backgeo}) and leads to
the useful identity that
\be
\bd_{b} \, \bd_{c} \, \bx^{a} = \frac{2 \Lambda}{(D-1)(D-2)} \,
(\d^{a}\,_{b} \, \bx_{c} - \eta_{bc} \, \bx^{a}) \, . \label{iden}
\ee

Consider, for example, the first term in (\ref{xitau}). One has 
\begin{eqnarray*}
\bx^{a} \, \bd_{c} \, \bd^{b} \, \vp^{c}\,_{b} & = & 
\bd_{c} \, (\bx^{a} \, \bd^{b} \, \vp^{c}\,_{b}) - 
\bd^{b} \, (\vp^{c}\,_{b} \, \bd_{c} \, \bx^{a}) +
\vp^{c}\,_{b} \, (\bd^{b} \, \bd_{c} \, \bx^{a}) \\
& = & \bd_{c} \, (\bx^{a} \, \bd^{b} \, \vp^{c}\,_{b} -
\vp^{bc} \, \bd_{b} \, \bx^{a}) + \frac{2 \Lambda}{(D-1)(D-2)} \,
(\vp_{c}\,^{a} \, \bx^{c} - \vp^{c}\,_{c} \, \bx^{a}) \, .
\end{eqnarray*}
We have carried the $\bx^{a}$ term `inside' the derivative operator in the
first line and used (\ref{iden}) to obtain the second line. One follows
similar steps for the other terms in (\ref{xitau}), and noting that all
terms of the type $\vp^{ab} \, \bx_{b}$ and $\vp^{b}\,_{b} \, \bx^{a}$
cancel out separately along the way, the final expression for (\ref{charden})
is found.

\begin{acknowledgments}

This work is partially supported by the Scientific and Technical Research
Council of Turkey (T{\"U}B\.{I}TAK). B.T. is also partially supported by
the ``Young Investigator Fellowship" of the Turkish Academy of Sciences 
(T{\"U}BA) and by the  T{\"U}B\.{I}TAK Kariyer Grant 104T177. H.C. 
acknowledges the support through a Post-Doctoral Research Fellowship by 
T{\"U}B\.{I}TAK.

\end{acknowledgments}

\end{document}